\documentstyle[bezier,epsfig,12pt,preprint,tighten,aps]{revtex}
\begin{document}
\draft

\title{\rightline{{\tt {March 1999}}}
\rightline{{\tt KIAS-P99028}}
\rightline{{\tt UDELHEP-99/102}}
\ \\
Mapping Lorentz Invariance Violations into Equivalence Principle Violations}
\author{A. Halprin$^{a,b}$\footnote{halprin@udel.edu}
and H. B. Kim$^a$\footnote{hbkim@muon.kaist.ac.kr}}
\address{$^a$Korean Institute for Advanced Study\\
207-43 Cheongryangri-dong, Dongdaemun-gu, Seoul 130-012, Korea\\
$^b$Department of Physics and Astronomy, University of Delaware\\
Newark, DE 19716, USA\\
}
\maketitle
\begin{abstract}
We point out that equivalence principle violations, while not dynamically equivalent, produce 
 the same kinematical  
effects as Lorentz invariance violations for particle processes in a
constant gravitational potential. This allows us to translate many
experimental bounds on Lorentz invariance violations into 
bounds on equivalence principle violations. The most stringent bound suggests that a postive
 signal in an E\"otv\"os experiment may be at least seven orders of magnitude beyond current
 technology.
\end{abstract}
   
 \newpage

Recently, new limits have been placed on possible violations of Lorentz 
invariance (LIV)\cite {cg1}, and it has been noted that such a violation
can turn on (turn off) processes at sufficiently high energy
that are  otherwise kinematically forbidden (allowed) \cite{cg1,gm}.
In particular, it has been suggested that such a violation may provide a
satisfactory answer to the observation of cosmic ray protons at energies
above the GZK cutoff \cite{gr,zk}.
In this paper we demonstrate that all effects of the type of
special relativity violation entertained by Coleman and Glashow can,
for all practical purposes in our solar system,
be obtained by the less rash hypothesis of small violations of
the equivalence principle (EPV) through a lack of universality
in gravitational coupling strengths to matter.
This has previously been noted in the case of neutrino oscillations
\cite{ghklp} but not for other phenomena. 
With this equivalence established, we use the same phenomena to place
limits on gravitational couplings to ordinary matter and set goals
for future E\"otv\"os type experiments.

The proposal of Coleman and Glashow \cite{cg1,cg2} is to modify the the
usual free particle dispersion relation of special relativity in a minimal
way by allowing the world line of each species
to be determined by its own limiting velocity , i.e.
\begin{equation}
E^2=(pc_a)^2+(mc_a^2)^2
\end{equation}
The limits on values of $c_a$ that have been obtained as well as
suppression of the GZK bound arise 
from the fact that in this scheme some particle velocities may exceed 
the limiting velocities of other particles. 

The same kind of ``superluminal'' effects happen if universality 
of the gravitational coupling is relaxed without giving up special relativity.
This is because current observations are made in a medium containing
an essentially constant gravitational potential, which does nothing more than
change the limiting speeds of particles in the medium in a species
dependent manner\cite{g3}.
Since we are concerned only with kinematics, we shall treat all particles
as spinless and therefore described by a Klein-Gordan equation.
The effect of gravity is to modify the metric from flat space time,
$\eta_{\alpha\beta}$, to the metric $\eta_{\alpha\beta} + h_{\alpha\beta}$.
In the weak field approximation to Einstein gravity written
in the harmonic gauge, $h_{\alpha\beta}$ is determined by \cite{w} 
\begin{equation}
\partial_{\alpha}\partial^{\alpha}h_{\mu\nu}=-16\pi GS_{\mu\nu}
\end{equation} 
where G is Newton's constant and the tensor source current,
$S_{\mu\nu}$, is related to the energy momentum tensor, $T_{\mu\nu}$, by
\begin{equation}
S_{\mu\nu}=T_{\mu\nu}-\frac{1}{2}\eta_{\mu\nu}T^{\lambda}_{\lambda}
\end{equation}
For a static matter source distribution producing Newtonian potential $\Phi$,
the solution for $h_{\mu\nu}$ is \cite{will}
\begin{equation}
h_{\mu\nu}=2\Phi\delta_{\mu\nu}
\end{equation}    

We introduce a breakdown of Einstein gravity by allowing
the coupling of the metric to the  Klein-Gordan equation
for particle species $a$ to be nonuniversal through a parameter $f_a$,
which is unity in Einstein theory, i.e., we make the replacement
\begin{equation}
h_{\mu\nu} \rightarrow f_a h_{\mu\nu}
\end{equation}
In the case of a constant Newtonian potential,
the otherwise free particle energy-momentum relation thus becomes
\begin{equation}
(1+2f_a\Phi)E^2=(1-2f_a\Phi)p^2 + m_o^2
\end{equation}
where $m_o$ is the mass in a true vacuum.
If all $f_a$ were equal, this constant potential would have no physical
significance, because it would just be absorbed in a redefinition of the
velocity of light. However, if two particles have unequal values of $f_a$,
a constant value of the Newtonian potential acquires physical significance,
since it means that their limiting velocities will differ. 
 
At some level, the physical significance of the absolute value of
the Newtonian potential will have to show up in a modification of Eq.~(2),
since it is invariant under $\Phi\rightarrow\Phi+{\rm constant}$. 
We would anticipate that the EPV portions of the coupling will contribute
to the graviton self energy to produce a small but non-zero graviton mass.
Of course, this is a matter of conjecture, since we have absolutely no way
to calculate such an effect.
The reader may wonder if such a graviton mass is ruled out by the
observation of Van Dam and Veltman \cite{vdv}
that a graviton with any mass is inconsistent with the experimental
accuracy on observations of the advance of the perihelion of Mercury.
We would argue that such an inference is not possible,
because generating a graviton mass in this way could very well result
in a smooth transition from the massive to the massless graviton similar,
for example, to that devised by Schwinger \cite{s}.

In order to cast the LIV and EPV dispersion relations in the same form,
we simply note that in the latter case it is natural to redefine the mass
such that in the low energy limit the kinetic energy is still related
to the observed mass $m$ by the usual nonrelativistic relation
$K.E.=m^2/2E$, in which case we have
\begin{equation}
m^2=\frac{(1+2f_a\Phi)}{(1-2f_a\Phi)^2}m_o^2
\end{equation}
and
\begin{equation}
E^2=(pc_a)^2+(mc_a^2)^2
\end{equation}
where $c_a$ and $f_a$ are related by 
\begin{equation}
c_a = \frac{(1-2f_a\Phi)}{(1+2f_a\Phi)} \simeq 1-4f_a\Phi.
\end{equation}

The kinematical equivalence of LIV and EPV is now manifest,
and we are now in a position to translate all of the limits
on differences of various $c_a$ compiled by Coleman and Glashow (CG)
into limits on differences of $f_a$ values once we have a value for
the Newtonian potential.
{}From the compilation of \cite{hlp}, we see that for experiments in our
galaxy the dominant contribution observed thus arises from the great
attractor and has a value of $\Phi=-3\times10^{-5}$ \cite{k}.
Assuming that any induced graviton mass as discussed above is irrelevant
at this distance scale, we use this value as a lower limit
(there may be other bodies out there that give an even larger contribution) 
for $\vert\Phi\vert$.
In Table~1 we give the CG limits on limiting velocities
and the corresponding limits on the gravitation parameters.

It should be emphasized that while the LIV and EPV scenarios are
kinematically indistinguishable,
they can be distinguished by the dynamics of low energy experiments,
because there is no trace of LIV in the low energy form of the kinetic energy.
For example, in an E\"otv\"os experiment, even if LIV is present,
there will be no positive signal unless the equivalence principle is violated.
The limit on $f_p$ obtained from ultra-high energy cosmic rays suggests that
observation of a positive signal in any E\"otv\"os experiment may be at least
some seven orders of magnitude away from current technology.

\begin{table}
\begin{tabular}{|c|c|c|} \hline
Related experiment & Bound on LIV & Bound on EPV \\\hline\hline
\parbox{60mm}{observation of ultra high energy cosmic rays
\cite{cg2}} &
$c_p-c_\gamma<1\times10^{-23}$ &
$f_p-f_\gamma<1\times10^{-19}$ \\\hline
\parbox{60mm}{high precision spectroscopic experiments that fail to find
anisotropies in CMB
\cite{ljhrf}} &
$|c_m-c_\gamma|<6\times10^{-22}$ &
$|f_m-f_\gamma|<6\times10^{-18}$ \\\hline
\parbox{60mm}{search for high energy neutrino oscillation
\cite{b}} &
$|c'-c|_{\nu_e\nu_\mu}<6\times10^{-22}$ &
$|f'-f|_{\nu_e\nu_\mu}<6\times10^{-18}$ \\\hline
\parbox{60mm}{muon fluxes in the air shower
\cite{cs}} &
$|c'-c|_{\mu e}<4\times10^{-21}$ &
$|f'-f|_{\mu e}<4\times10^{-17}$ \\\hline
\parbox{60mm}{energy dependence of $K_L$-$K_S$ mass difference
\cite{hms}} &
$|c_{K_L}-c_{K_S}|<3\times10^{-21}$ &
$|f_{K_L}-f_{K_S}|<3\times10^{-17}$ \\\hline
\end{tabular}
\caption{Bounds on the equivalence principle violation parameters
mapped from bounds on the Lorentz invariance violation parameters}
\end{table}

\end{document}